\documentclass[authoryear]{elsarticle}

\usepackage[top=1.25in, bottom=1.25in, left=1.5in, right=1.5in]{geometry}
\usepackage{lineno}
\usepackage{soul}
\usepackage{color}
\usepackage{tikz}
\usetikzlibrary{quotes,angles}
\usepackage[pagebackref=true,
            colorlinks=true,
            bookmarks=true,
           ]{hyperref}   
           
\usepackage[most]{tcolorbox}           
\usepackage{graphicx}
\usepackage{epstopdf}
\DeclareGraphicsExtensions{.eps}
\usepackage{siunitx}

\usepackage{empheq}
\usepackage{amssymb}
\usepackage{array}
\usepackage{amsmath, scalerel}
\usepackage{leftidx}

\usepackage{amssymb}
\usepackage{gensymb}

\usepackage{amsthm}

\usepackage{mathptmx}

\usepackage{amsfonts}
\usepackage{multicol}
\usepackage{mathrsfs}
\usepackage{tensor}
\usepackage{subfig}
\usepackage{hhline}
\usepackage{upgreek}
\usepackage{cancel}
\usepackage{ulem}
\newcommand*\circled[1]{\tikz[baseline=(char.base)]{
            \node[shape=circle,draw,inner sep=1pt] (char) {#1};}}

\usepackage{multirow}
\usepackage{setspace}

\usepackage[flushleft]{threeparttable}
\usepackage{makecell,booktabs}

\usepackage{amsmath,scalerel}

\newcommand{\tightoverset}[2]{\mathop{#2}\limits^{\vbox to -.5ex{\kern-0.75ex\hbox{$#1$}\vss}}}
\modulolinenumbers[5]

\journal{Elsevier}

\begin{document}
\nolinenumbers

\begin{frontmatter}
\title{Extensive anisotropic lath martensite plasticity in dual-phase steels: A numerical-experimental investigation}

\author[1,2]{V. Rezazadeh}
\author[1]{\corref{cor}R.H.J. Peerlings}\ead{r.h.j.peerlings@tue.nl}
\author[1,2]{T. Vermeij}
\author[1]{J.P.M. Hoefnagels}
\author[3]{F. Maresca}
\author[1]{M.G.D. Geers}

\address[1]{Department of Mechanical Engineering, Eindhoven University of Technology (TU/e), P.O.Box 513, 5600 MB Eindhoven, The Netherlands}
\address[2]{Materials Innovation Institute (M2i), P.O.Box 5008, 2600 GA Delft, The Netherlands}
\address[3]{Faculty of Science and Engineering, P.O.Box 72, 9747 AG  Groningen, The Netherlands}

\cortext[cor]{Corresponding author.}

\begin{abstract}
This work presents a detailed experimental-numerical analysis of a low-carbon dual-phase steel microstructure, revealing evidence of significant anisotropic plastic deformation in lath martensite.    
A careful determination of the habit plane orientations in the present martensitic variants demonstrates that the observed traces of plastic slip coincide with the directions of the corresponding habit planes.   
To study the local lath martensite plastic activity, a dedicated substructure-enriched crystal plasticity based model is exploited. 
Compared with conventional bcc crystal plasticity, the model incorporates an extra crystallographic slip plane containing 3 softer slip systems parallel to the habit planes to capture the role of habit-plane plasticity.
Simulations reveal consistent strain localization patterns as those observed in the experiments, with most plasticity localized in the martensite packets with their habit plane oriented favorably with respect to the applied load.
Based on these insights, recommendations for future martensite modelling strategies and potential improvements of steels are discussed.

\end{abstract}

\end{frontmatter}

\section{Introduction}
	Lath martensite, as the most common morphology of martensite in low-alloyed steels, is the primary strong phase in a large variety of commercial advanced high strength steels. For instance, in dual-phase (DP) steels an optimal balance of ductility and strength is obtained by pairing ferrite and martensite phases in their microstructure. Several strengthening mechanisms explain the high base-strength of lath martensite, including solid solution hardening by alloying elements \citep{KRAUSS199940,Takaki2012}, carbon Cottrell atmosphere effects \citep{wilde2000Cottrell}, carbide precipitation hardening \citep{KRAUSS199940, OHMURA20031157}, and forest dislocation hardening \citep{Takaki2012, azevedo1978mossbauer}. Furthermore, an abundance of internal boundaries originating from its complex hierarchical substructure, including packets, blocks, and sub-blocks, contribute significantly to the high strength of lath martensite \citep{bhadeshia2017steels, MORITO2006237, MINE2013535, DU2016117}. However, it has been confirmed that under a favorable orientation of the applied load some of these internal boundaries may also accommodate a significant amount of plastic deformation. In particular, experimental evidence of lath interface plasticity has been reported in multiple studies, for instance, by observing extreme slip steps appearing along the lath boundaries in micro-tension and micro-compression tests performed on single block/packet martensitic specimens extracted from fully martensitic and DP steels \citep{ghassemi2013microscale, MINE2013535,du2016plasticity,kwak2016anisotropy,du2019lath,ghaffarian2022interfacial,TIAN2020274, LIU2021116533}. Plastic localization traces parallel to the substructure boundaries have also been observed in mesoscale experiments done on low-carbon martensitic steel \citep{inoue2019slip,morsdorf2016multiple}.

	The apparent ductility of lath martensite, in directions parallel to the lath boundaries, triggered a renewed interest in the detailed understanding of its underlying deformation mechanisms. In this regard, three dominant hypotheses have been suggested in the literature: 
\begin{enumerate}
\item A number of studies have found small fractions of the inter-lath retained austenite in martensitic steels and dual phase steels \citep{SANDVIK1983199,SAMUEL198551,morito2011carbon,COTAARAUJO2021106445,yoshida2015crystallographic, liao2010microstructures}. 
	Using crystal plasticity simulations, \citet{Maresca2014} have demonstrated that such fcc films may act as 'greasy films' along which the harder laths can slide. 
\item Others have suggested that phase transformation of the inter-lath retained austenite may contribute significantly to the apparent plasticity in lath martensite \citep{morito2011carbon, morito2007comparison, morsdorf2016multiple, MARESCA2017302, MARESCA2018463, inoue2019slip}. 
\item Due to the characteristic lath morphology, active dislocations on the intra-lath $\{110\}\langle111\rangle$ bcc systems have long  mean free paths. As a result, slip in these directions can proceed comparatively freely \citep{MICHIUCHI20095283, MINE2013535, harjo2017work, ungar2017composite}.		
\end{enumerate}	

	So far, the extreme plastic localization parallel to the lath boundaries has been observed predominantly in fully martensitic steels with a coarse substructure, obtained by dedicated heat-treatments and a low carbon content. Such microstructures facilitate visualization at small scales, yet are not representative for dual-phase steels, since there is no surrounding ferrite, which is likely to affect the martensite plasticity. Our objective in this study is twofold. First, we will lift the limitation to single-phase martensitic materials by examination of the lath martensite response in-situ in a dual-phase steel microstructure. Furthermore, to carefully unravel the underlying mechanisms of the experimental observations, a one-to-one numerical-experimental coupled approach will be employed, whereby the microstructural characterization of DP steel grains/phases and their deformation behavior is compared with simulations performed by using the same experimental morphology and boundary conditions, by means of a novel substructure-informed CP-based numerical model.
	
\section{Experimental investigation}
	Lath martensite in commercial dual-phase steels typically exhibits a significantly distorted and dislocated substructure, due to its higher carbon content, inhibiting proper microstructural and micro-mechanical characterization of the lath boundaries \citep{DU2018411, du2019lath, TIAN2020274}. These limitations can be tackled by producing a dual-phase steel microstructure consisting of coarse lath martensite islands surrounded by soft ferrite grains, thereby enabling advanced characterization of the lath morphology and corresponding deformations \citep{Vermij2022, LIU2021116533}. 

\begin{figure}[ht!]
\centering
  \includegraphics[width=1\linewidth]{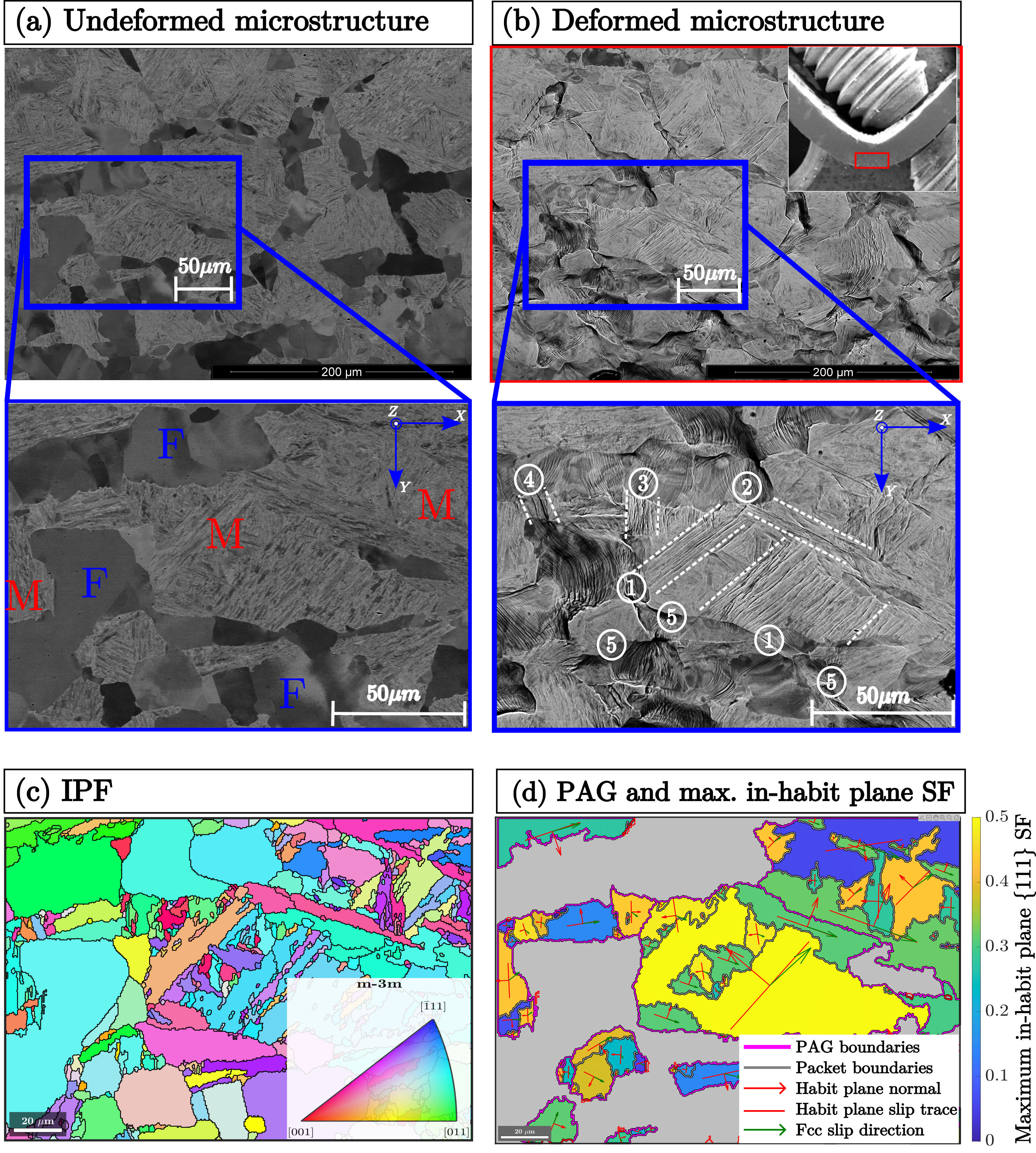}
  \caption{SEM image in BSE mode of the (a) undeformed, and (b) deformed configuration of the heat treated dual-phase steel microstructure. The dark phase is ferrite and the brighter phase is martensite. The enlarged areas show the region of interest (ROI) considered for this study. The white dashed lines tagged \textcircled{1} to \textcircled{4} indicate the regions in which plastic slip occurs parallel to the habit plane traces. The regions tagged \textcircled{5} show limited plastic deformation. (c) The inverse pole figure (IPF) map of the ferrite grains and martensite variants. (d) The reconstructed prior austenite grain (PAG) boundaries (\textcolor{violet}{\rule{5ex}{2pt}}), packet boundaries (\textcolor{gray}{\rule{5ex}{1pt}}), habit plane normal (\textcolor{red}{$\longrightarrow$}) and the habit plane trace of each packet (\textcolor{red}{\rule{5ex}{1pt}}), and in-habit plane ($\{111\}$) fcc slip system direction with the highest Schmid factor (SF) (\textcolor{green}{$\longrightarrow$}). The color code indicates the computed maximum in-habit plane SF based on an applied tensile load in the horizontal direction.}
  \label{fig:experiments}
\end{figure}

A commercial DP600 grade provided by Tata Steel Europe, with composition of 0.092C-1.68Mn-0.24Si-0.57Cr wt.$\%$ was heat-treated as follows: 10 minutes austenization at 1100°C, followed by a 30 minutes inter-critical anneal at 750°C, and water quenching to room temperature \citep{Vermij2022}.	
	A clean DP microstructure was thus obtained, with martensite island sizes in a range of $50\mu m$ to $100 \mu m$, and a martensite volume fraction of $\approx60\%$. 
	Figure \ref{fig:experiments}(a) shows a Scanning Electron Microscopy (SEM) image of the microstructure obtained in Back-Scattered Electron (BSE) mode. The lighter phase is martensite and its substructure is visible in detail. The enlarged area is chosen as the region of interest (ROI) in what follows. 	
	
	A sample with an approximate size of $30\times 8 \times 1 \, \mathrm{mm}^{3}$ was cut, mechanically polished and tested in a tight angle hemming test set-up. The main advantage of the hemming test is that it enables the application of large strains without formation of a localized neck. Figure \ref{fig:experiments}(b) shows the dual-phase microstructure of Figure \ref{fig:experiments}(a) at $\approx 30\%$ of tensile strain in the horizontal direction. The deformed image is re-scaled in order to span the same field of view as the undeformed image. The images shown in Figure \ref{fig:experiments}(a \& b) were taken on the surface of the sheet which is normal to the bending axis, close to the outer outer surface, i.e. in the region with the highest tensile strains.

	The enlarged region in Figure \ref{fig:experiments}(b) shows significant localized slip bands which have occurred in particular regions of certain martensite islands, e.g. shown in the enlarged area by overlaid white dashed lines tagged \circled{1} to \circled{4}. On the contrary, some regions inside and in the vicinity of the same island do not show any clear slip activity, cf. the regions tagged \circled{5}.	

	Before deforming the specimen, the crystallographic orientations of both phases were mapped with Electron BackScatter Diffraction (EBSD) employing spherical indexing for robust identification of martensite orientations \citep{LENTHE2019112841}. A misorientation threshold of $2.5^\circ$ was used to identify the individual variants of the martensite \citep{MORITO20065323}. Figure \ref{fig:experiments}(c) shows the Inverse Pole Figure (IPF) map of the ferrite grains and martensite variants in the ROI, which were distinguished through thresholding of the EBSD image quality. By using the crystallographic orientation of each variant, and by taking into account the Kurdjumov-Sachs (KS) orientation relationship $\{111\}_{\gamma} \, || \, \{011\}_{\alpha^{\prime}}$ and $\langle 101 \rangle_{\gamma} || \langle 111 \rangle_{\alpha^\prime}$ \citep{Kurdjumow1930}, the Prior Austenite Grains (PAG), as well as the martensite packets, blocks and sub-blocks within PAG were reconstructed, using the PAG reconstruction tools from the Matlab MTex Toolbox \citep{Hielschercg5083, HIELSCHER2022101399, niessen2022parent,nyyssonen2016iterative}. The resulting habit plane traces and maximum Schmid Factor (SF) of in-habit plane fcc slip systems of the potential inter-lath austenite films were identified for each packet and presented in Figure \ref{fig:experiments}(d).		  
	
	By comparing the slip traces observed in the deformed ROI, shown in Figure \ref{fig:experiments}(b), with the habit planes traces, shown in Figure \ref{fig:experiments}(d), we observe that the plastic slip, in the regions highlighted by white dashed lines tagged \circled{1} to \circled{4}, occurs on directions parallel to the habit planes and hence to inter-lath boundaries. It is further shown that martensite packets with more unfavorable habit plane orientations, i.e. low maximum Schmid factor (SF), do not form slip traces, which indicates limited plastic deformation, see the regions tagged \circled{5} in Figure \ref{fig:experiments}(b).

\section{Numerical investigation}
\begin{figure}[ht!]
\centering
  \includegraphics[width=0.8\linewidth]{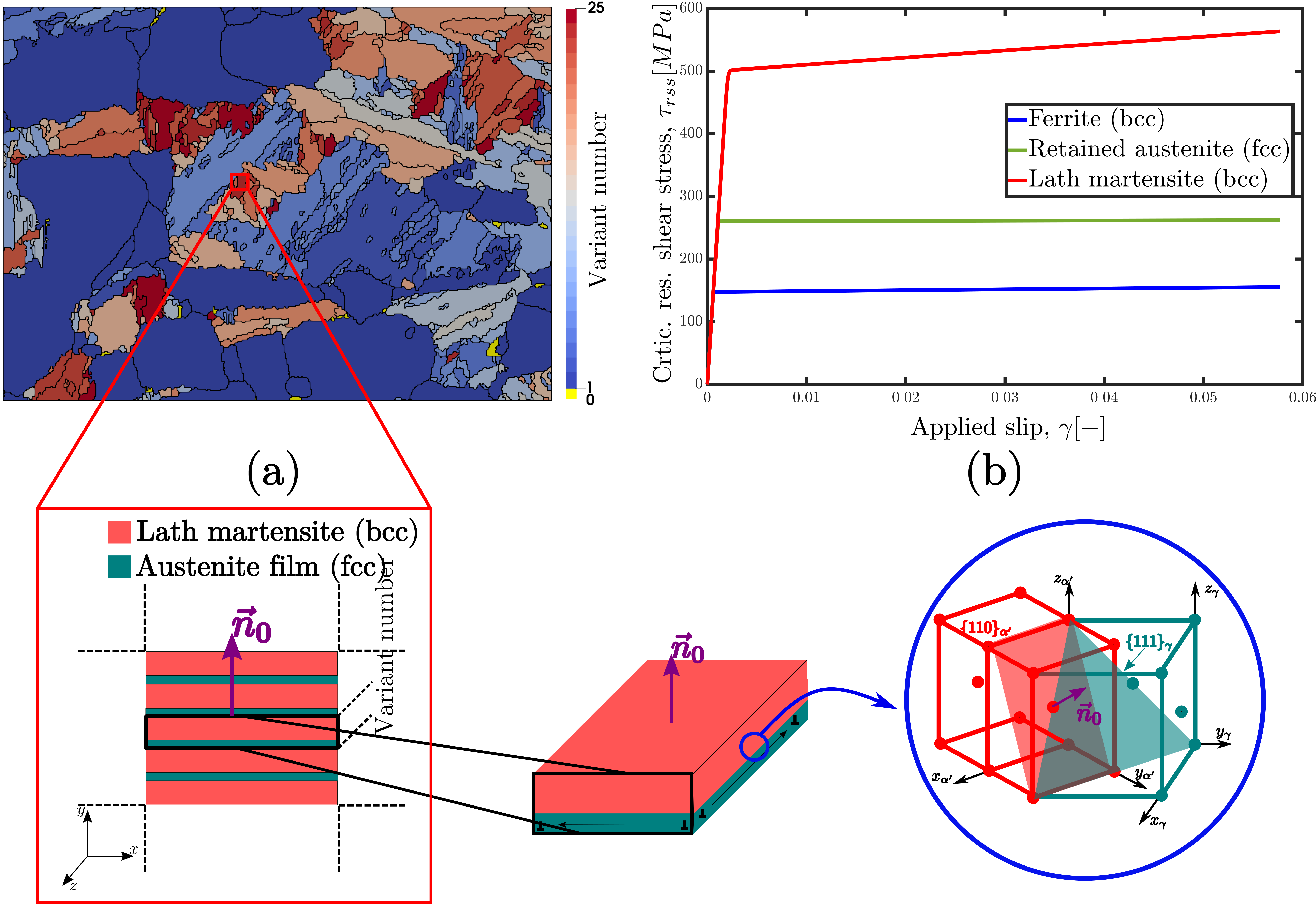}
  \caption{a) The two-dimensional (2D) microstructural model of the ROI. The ferrite grains are indicated by $ID=1$ on the color bar, while the martensite variants are numbered from $ID=2-25$; and the $ID=0$ is for unindexed martensite variants which were not assigned to any PAG by the algorithm. For the unindexed variants conventional CP model with lath martensite (bcc) properties is used. b) Stress-strain response of a single slip system of different phases used in the simulations \citep{MARESCA2016198}. }
  \label{fig:Modelling}
\end{figure}

	To further rationalize the above experimental observations, we first construct a two-dimensional (2D) numerical model of the region of interest shown in Figure \ref{fig:experiments}(a) -- see Figure \ref{fig:Modelling}(a). The model is based on conventional crystal plasticity (CP) for ferrite and a dedicated, substructure-enriched martensite CP model which accounts for the additional, softer plastic deformation mechanisms in the inter-lath retained austenite films -- or parallel to them. The enrichment provides, next to the $12$ or $24$ common bcc slip systems of the martensite laths, 3 fcc slip systems representing the austenite film and are, according to the KS orientation relationships, parallel to the film and hence to the habit plane. These slip systems are assigned properties which are representative for austenite, and they hence are softer than the conventional bcc systems, which have properties representative of the martensite laths (see Figure \ref{fig:Modelling}(b)); properties taken from \citet{maresca2014subgrain}. Note that this model is based on the hypothesis that the plastic slip of the inter-lath retained austenite films is the main carrier of martensite yielding, i.e. hypothesis $1$. It is however believed to also capture, albeit in a more qualitative manner, the mechanical effect of the other two candidate hypotheses listed above. The detailed formulation of the model is give in \ref{sec:Ch6-ConstRelations}. While revealing a similar mechanical response, the model is simpler and more efficient than the reduced martensite model of \citet{MARESCA2016198}. Predictions made with this model are compared below in Figures \ref{fig:Result1} with those using solely conventional CP for martensite. Horizontal straining is applied via periodic boundary conditions for a uniaxial mean stress state, i.e. free contraction in vertical and out of plane directions. All simulations were done in the DAMASK simulation package \citep{ROTERS2019420}. 

	Figure \ref{fig:Result1} shows the predicted deformation in the 2D ROI obtained using the enriched and conventional CP models. The response of the ferrite regions is not shown/discussed here as it is not the focus of this study. The results are given based on the 2D equivalent true (logarithmic) strain, $\varepsilon$, computed in the model plane, and the ratio of the in-habit plane (IHP) slip, ($\gamma_{\mathrm{IHP}}$), to the total slip of each variant, $\gamma_{\mathrm{total}}= \gamma_{\mathrm{IHP}} + \gamma_{\mathrm{OHP}}$, in which $\gamma_{\mathrm{OHP}}$ is the total slip in out-of-habit plane slip systems. This ratio characterizes the contribution of in-habit plane slip systems with respect to the overall plastic deformation of the crystal. In the enriched CP model, $\gamma_{\mathrm{IHP}}$ is governed mostly by the softer fcc slip systems; in the conventional CP (without fcc slip systems) $\gamma_{\mathrm{IHP}}$ containes only the slip on the two bcc systems parallel to the habit plane.
\begin{figure}[ht!]
\centering
  \includegraphics[width=1\linewidth]{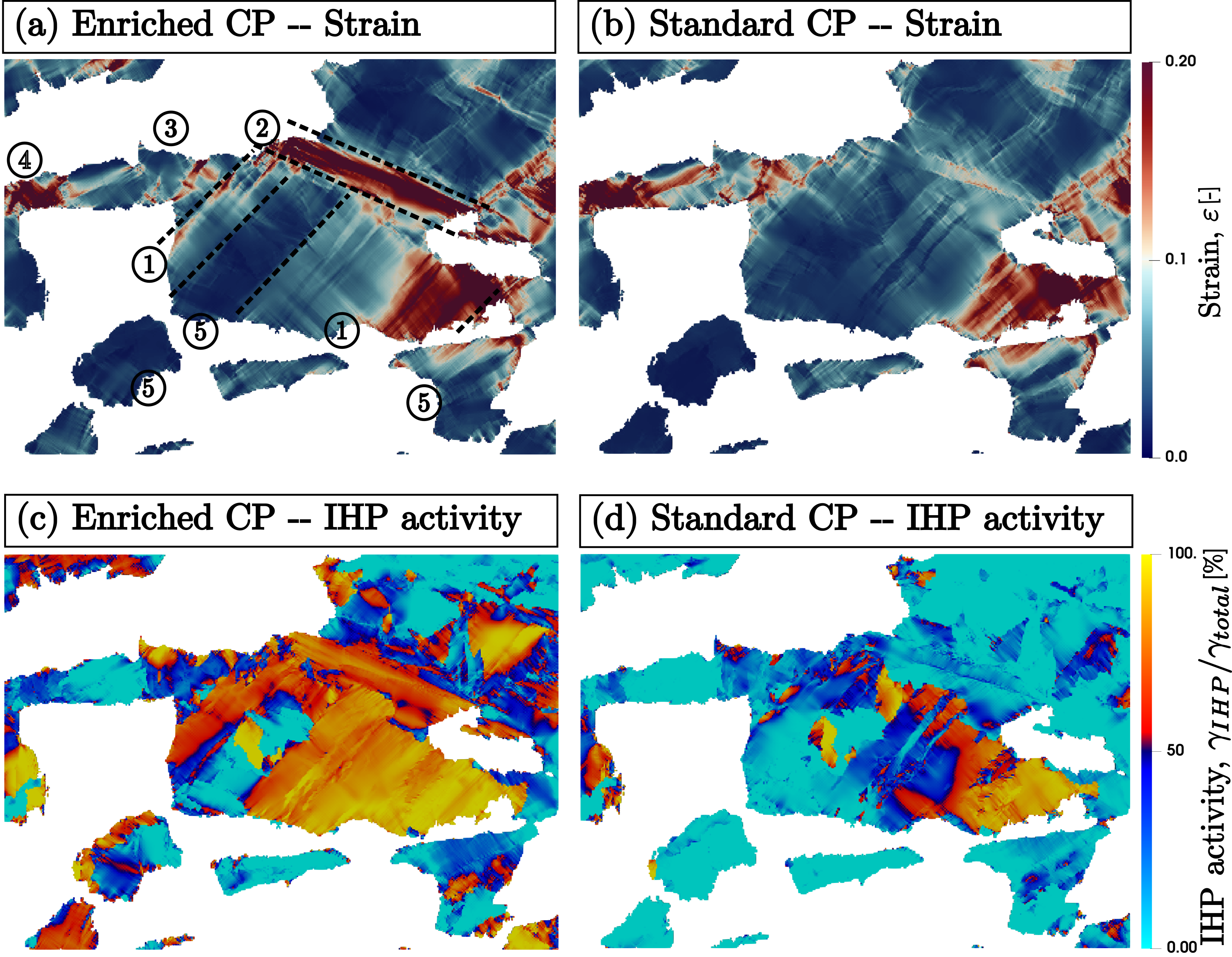}
  \caption{ a), b) Equivalent in-plane strain map, and c), d) in-habit plane (IHP) slip activity ratio map obtained for the martensite variants modeled with a \& c) the enriched CP model, b \& d) conventional CP model, respectively. The numbered areas are discussed in the text. }
  \label{fig:Result1}
\end{figure} 

	It is observed in the strain map, as shown in Figure \ref{fig:Result1}(a), that the deformation in the martensite islands is highly heterogeneous. The areas tagged \circled{1}--\circled{2} have a high degree of localized strain, whereas the areas tagged with \circled{5} show comparatively lower plastic activity. The IHP activity shown in Figure \ref{fig:Result1}(c), reveals that, in the highly localized regions, the IHP slip systems are dominantly active. These predictions are in good agreement with the slip activity observed in the experiments, cf. Figure \ref{fig:experiments}(b). It is noticeable that the model tends to accommodate the plastic strain by the additional fcc systems, even when their orientation is not highly favorable, i.e. for low in-habit plane SF. The region tagged \circled{2} is an example of this case. The model in which martensite variants are modeled with standard CP is not able to pick up the strain localization pattern observed in the experiments, see Figure \ref{fig:Result1}(b). This is shown in Figure \ref{fig:Result1}(d), where in all of the martensite variants that show much plasticity in the experiments the simulated OHP activity is more dominant.

\begin{figure}[ht!]
\centering
  \includegraphics[width=1\linewidth]{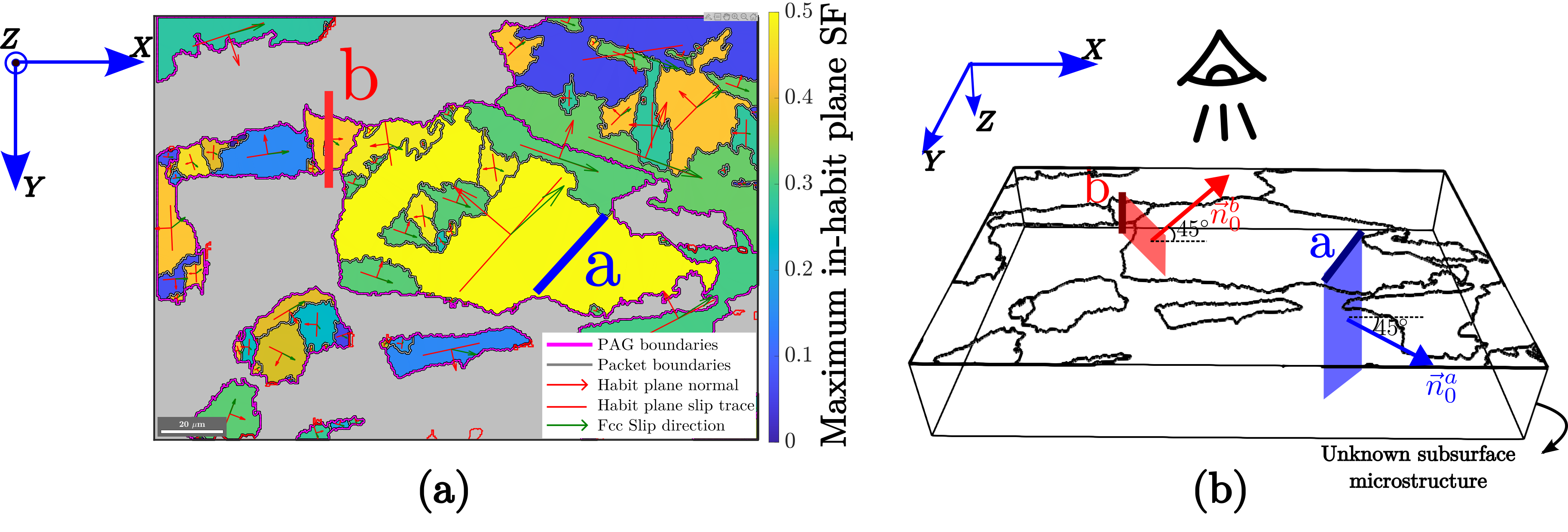}
  \caption{ a) The reconstructed packet map of the ROI, colored based on the computed maximum in-habit plane SF. The habit planes of two packets with $SF\approx0.5$ but different orientation with respect to the plane of the 2D model are indicated.
b) 3D sketch of the orientation of the two planes with respect to the sample surface. The 2D model assumes the grains/packets to be columnar. This may restrict the activity of plane $b$ while leaving $a$ unaffected. The real subsurface microstructure is unknown.}
  \label{fig:3DSubstructure}
\end{figure}

	Despite the capability of the model to predict most regions of plastic localization, i.e. in the areas tagged \circled{1} \& \circled{2}, a few inconsistencies with the experiment are also observed.  For instance, in the experiment, the areas tagged \circled{3} \& \circled{4} show significant slip lines parallel to the traces of the habit planes, see Figure \ref{fig:experiments}(b). However, the simulation results of IHP activity, shown in Figure \ref{fig:Result1}(c), indicate values well below $50\%$. By comparing the maximum SF map shown in Figure \ref{fig:experiments}(d) and the obtained numerical results shown in Figure \ref{fig:Result1}(a), we observe that in regions with a high in-habit plane SF, if the in-plane orientation of the corresponding habit plane (i.e. the orientation of the habit plane trace) is $\approx 45^\circ$, the responses are predicted correctly. However, deformation in regions with habit plane traces that are oriented vertically, even with similarly high in-habit plane SF, are not adequately captured by the model. This is presumed to be an artifact of the 2D modelling, which enforces uniformity of the microstructure in the out-of-plane direction, resulting in the obstruction of slip bands that intersect this columnar microstructure, see e.g. plane $b$ in Figure \ref{fig:3DSubstructure}, while slip bands parallel to it are unobstructed, see e.g. plane $a$ in Figure \ref{fig:3DSubstructure}. In reality, the subsurface microstructure is unlikely to be columnar and the out-of-plane slip planes are expected to be less obstructed.

\begin{figure}[ht!]
\centering
  \includegraphics[width=0.8\linewidth]{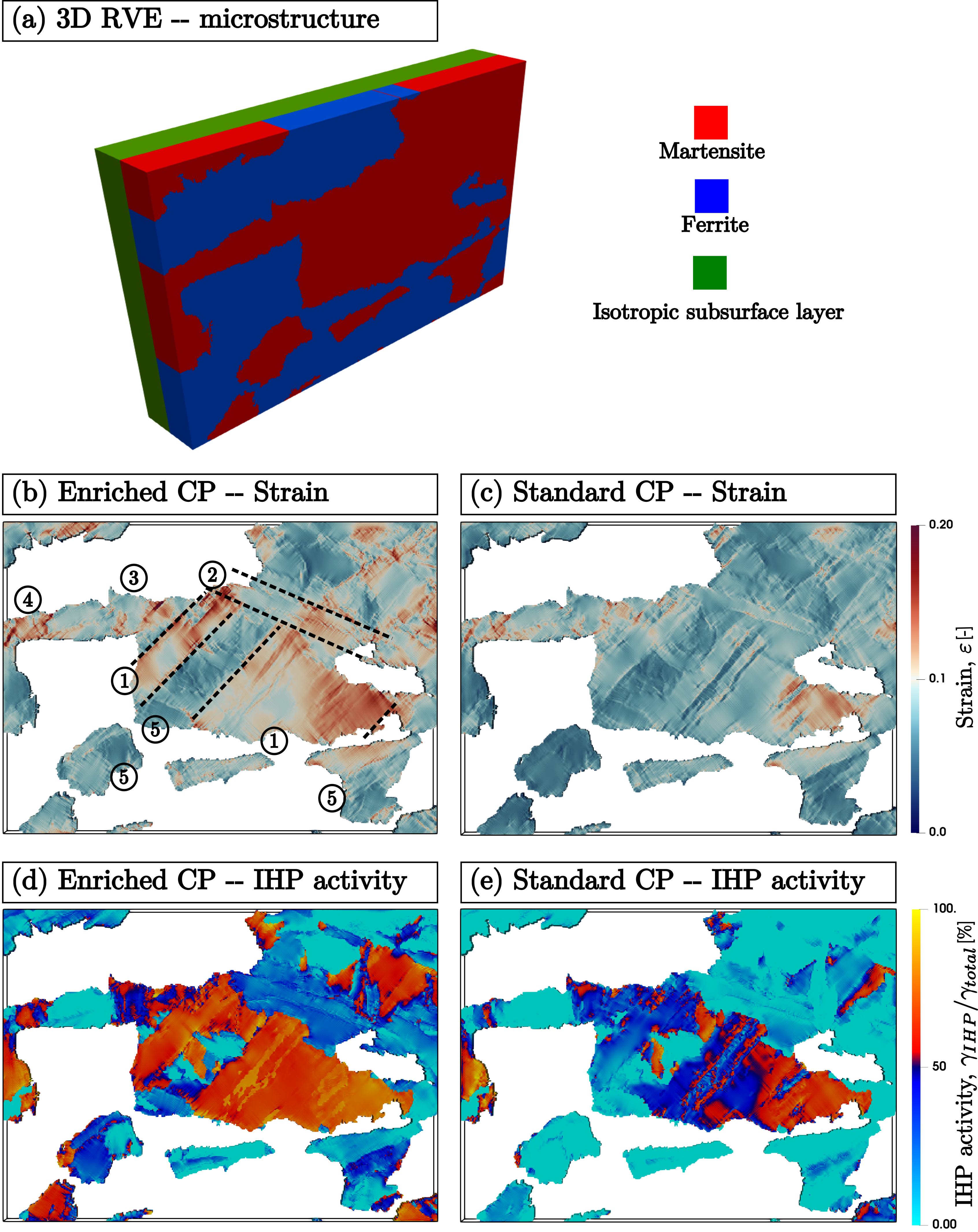}
  \caption{ a) 3D model with isotropic subsurface layer representing the unknown subsurface microstructure. b), c) The equivalent strain map, $\varepsilon$, for martensite variants modeled with the b) enriched CP, and c) standard CP model. The numbered areas are discussed in the text. d), e) The in-habit plane slip activity (IHP) ratio for the the martensite variants modeled with the d) enriched CP, and e) standard CP model.}
  \label{fig:Result2}
\end{figure}

To eliminate these artifacts, a 3D model has been constructed, by duplicating the 2D layer of the ROI microstructure in $z$-direction 10 times with the same pixel distance as in $x$ and $y$-direction, and then placing another 10 layers of an isotropic material with the combined properties of the ferrite and martensite. The effective properties of the isotropic substructure layer were computed by applying the rule of mixture of $60\%$ martensite and $40\%$ ferrite's slip resistance (of entire hardening curve), according to the volume fractions as measured in the ROI.  The single grain buffer layer represents the effective (unknown) substructure of the DP microstructure.
The resulting 3D microstructure is shown in Figure \ref{fig:Result2}(a).
 
Figure \ref{fig:Result2}(b \& c) shows the strain distribution in the martensite variants as computed using the enriched and the standard CP model. In both cases strains are observed to be more homogeneous, compared to the 2D models, cf. Figure \ref{fig:Result1}(a \& b). Since the out-of-plane microstructure is not constant anymore, more crystallographic slip planes can activate. For the same reason, in the 3D model, the IHP slip activity of the tagged regions \circled{3} \& \circled{4} (see Figure \ref{fig:Result2}d), has increased compared to the 2D case, Figure \ref{fig:Result1}(c), which is in accordance with the experimental observations (Figure \ref{fig:experiments}(b)). Furthermore, the highly localized region of the 2D model with an unfavorable habit plane orientation, tagged \circled{2} in Figure \ref{fig:Result2}(a), is less active in the 3D model simulation, yielding a much better overall agreement with the experiments. The comparison with the standard CP model indicates that the latter is unable to recover the patterns observed in the experiments. Compared with the 2D models the mismatch with the experiments has decreased -- mostly because the 3D enriched model captures the experimental maps much more accurately.

\section{Conclusions}
Based on a detailed experimental-numerical slip analysis, we conclude that, a soft plasticity mechanism is active along the habit plane in martensite islands of DP steels, with similar characteristics to what was previously observed only in fully martensitic steels.  
  The novel enriched CP model predictions reveals a good qualitative agreement with the plastic slip activity observed in the experiments. The standard CP model fails to predict many of the key activity locations seen experimentally. The only difference between the models are the three slip systems representing the intra-lath retained austenite films. Therefore, our results suggests that slip of these films, or at least a favorable plasticity-like mechanism parallel to these films (and hence to the habit plane) are essential contributors to the plastic deformation of lath martensite in the DP microstructure considered.  

Based on these insights, recommendations can be made for further research and optimization of DP steels. The soft habit plane plasticity mechanism requires implementation into future martensite models, to allow more accurate prediction of the behavior of newly designed steels, especially when local phenomena such as damage and edge cracking are of interest. More interestingly, it may be feasible to manipulate the orientation of the habit plane in critical damage-sensitive areas to promote plasticity, thereby avoiding local martensite brittle fracture.

\section*{Acknowledgements}

The authors thank R.L.A. Kerkhof and M. van Maris for experimental contributions and support. S. Roongta from MPIE D\"{u}sseldorf is gratefully acknowledged for the help in implementation of the model into the DAMASK software. The authors acknowledge F. Roters, and M. Diehl for the useful discussions on the numerical aspects. This research was carried out under project number T17019b in the framework of the research program of the Materials Innovation Institute (M2i) (\href{www.m2i.nl}{www.m2i.nl}) supported by the Dutch government.

\newpage

\appendix
\section{Constitutive relations of the proposed model}
\label{sec:Ch6-ConstRelations}
To capture the plastic localization parallel to the lath boundaries, we incorporate a modelling framework similar to \citep{VRezazadeh2022}. Each integration point (or voxel) of the martensite phase shown in Figure \ref{fig:Modelling2}(a) represents an infinite periodic two-phase laminate of a comparatively hard lath martensite and inter-lath retained austenite films, as shown in Figure \ref{fig:Modelling2}(b). Austenite films are infinitesimally thin that are considered as discrete planes embedded in the matrix of lath martensite. However, the model proposed in this paper is different from the one in \citep{VRezazadeh2022}, since it explicitly considers the crystallography, i.e. the orientations of slip systems, of lath martensite and austenite films instead of the isotropy assumption considered there.
Moreover, a certain orientation relationship is also taken into account between the two crystals at the interface plane, see Figure Figure \ref{fig:Modelling2}(d). 
The model can be described as a standard crystal plasticity formulation where a particular plane of a crystal, e.g. the habit plane, is enriched by 3 additional (softer) slip systems lying in that plane. While in-plane slip systems of retained austenite are presented by this softer crystallographic plane, the plasticity in the other 9 slip systems of the fcc crystal is constrained by the hard laths, as a result of the characteristic morphology of the lath martensite, and hence their contribution to plasticity is limited to what is already covered by the bcc crystal, see Figure \ref{fig:Modelling2}(c). 
Notably, even though we consider the austenite film sliding as the main plastic mechanism in the habit plane of lath martensite, the framework is applicable to two other hypotheses listed above.

\subsection{Elasticity}

	A standard crystal plasticity model is incorporated that is based on the phenomenological modelling of dislocation glide, resulting in plastic slip of two crystal regions with respect to each other along a slip plane \citep{peirce1983material, bronkhorst1992polycrystalline}. 
	Here, we employ a large strain formulation \citep{ANAND1985213}, where the deformation behavior of the model is characterized by the deformation gradient, $\mathbf{F}$, that can be multiplicatively split into an elastic contribution $\mathbf{F}_{\mathrm{e}}$, and an isochoric plastic contribution $\mathbf{F}_{\mathrm{p}}$, as follows,
\begin{align}
\mathbf{F}= \mathbf{F}_{\mathrm{e}} \cdot \mathbf{F}_{\mathrm{p}}.
\end{align}
	The elastic part incorporates the rigid-body rotation and the elastic strain.
	Due to the relatively small elastic deformations of a crystal, a Saint Venant-Kirchhoff type of model is adopted in which the induced elastic second Piola-Kirchhoff stress, $\mathbf{S}_\mathrm{e}$, is expressed as a function of the elastic Green-Lagrange strain tensor, $\mathbf{E}_e$, via the generalized \textsc{Hooke}'s law,
\begin{align}
\mathbf{S}_\mathrm{e}= \mathbb{C}:\mathbf{E}_\mathrm{e}.
\end{align}
	Here, $\mathbb{C}$ denotes the 4th-order elasticity tensor, inheriting the symmetry of a crystal lattice.  
	The model is elastically homogeneous, i.e., lath martensite behavior determines the elastic response, as it is intuitive that boundary sliding mechanism is in nature an extreme plastic deformation.

\subsection{Plasticity}

\begin{figure}[ht!]
\centering
  \includegraphics[width=1\linewidth]{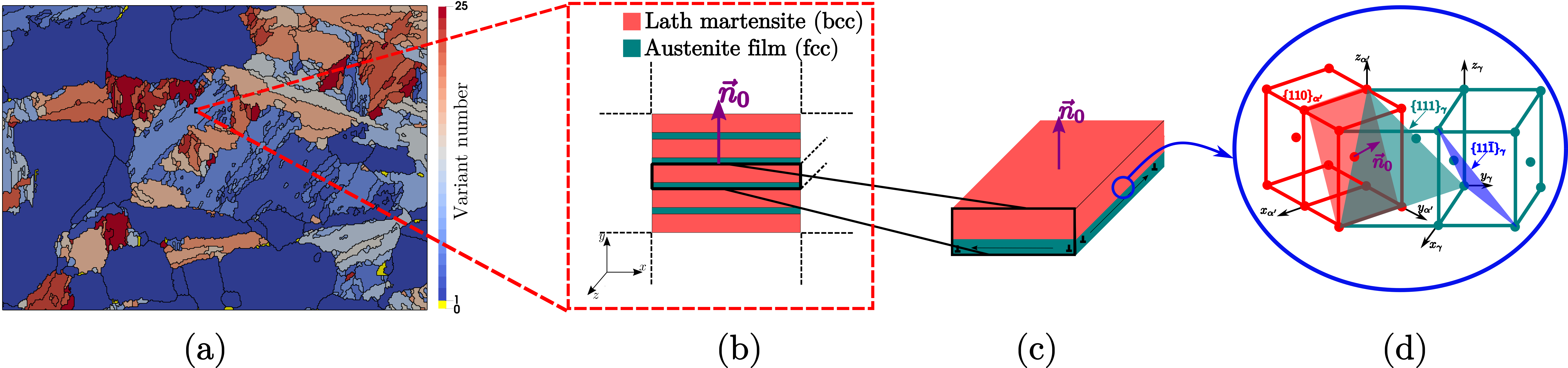}
  \caption{a) The two-dimensional (2D) microstructural model of the ROI as shown in Figure \ref{fig:Modelling}. b) Each voxel represents an infinite periodic laminate of martensite laths with embedded retained austenite films. c) A unit cell of martensite lath and austenite film illustrating that dislocations can only glide in the in-plane directions. d) KS orientation relationship indicates that, e.g. for the first packet the planes $\{111\}_{\gamma} \, || \, \{011\}_{\alpha^{\prime}}$ are parallel \citep{Kurdjumow1930}.}
  \label{fig:Modelling2}
\end{figure}

 Plasticity in lath martensite and austenite films occurs on the crystallographic slip systems of the corresponding bcc and fcc lattices. 
 The plastic velocity gradient $\mathbf{L}_{\mathrm{p}}$ is thus calculated as the sum of all the individual contributions via the rule of mixture,
\begin{align}
\label{eq:Ch6-Lpoverall}
\mathbf{L}_{\mathrm{p}} = \varphi \mathbf{L}_{\mathrm{p},{\gamma}} + (1-\varphi) \mathbf{L}_{\mathrm{p}, {\alpha^{\prime}}}. 
\end{align}
  with $\varphi$ being the volume fraction of retained austenite in lath boundaries. 
  The plasticity of retained austenite, generally speaking boundary sliding, can be decomposed into two separate contributions of the in-habit plane (IHP) and out-of-habit plane (OHP), 
\begin{align}
\mathbf{L}_{\mathrm{p},{\gamma}} = \mathbf{L}_{\mathrm{p}, \gamma} ^{\text{IHP}} + \mathbf{L}_{\mathrm{p},{\gamma}}^ {\text{OHP}}. 
\end{align}
An example of IHP and OHP systems are shown in Figure \ref{fig:Modelling2}(d) by a green $\{111\}_{\gamma}$ plane, which shares the $\{110\}_{\alpha^{\prime}}$ habit plane of the bcc crystal, and blue $\{11\bar{1}\}_{\gamma}$ plane of a fcc crystal, respectively. 
However, due to the characteristic morphology of lath martensite, i.e. elongated laths in the direction of the habit plane, the out-of-plane plasticity of austenite is controlled by the plasticity of lath martensite. This is schematically illustrated in Figure \ref{fig:Modelling}(c), by dislocations gliding only in the in-habit plane directions of the austenite phase.  
Therefore, we can assume $\mathbf{L}_{\mathrm{p},{\gamma}}^ {\text{OHP}}=\mathbf{L}_{\mathrm{p}, {\alpha^{\prime}}}$. 
This implies that Eqn. \ref{eq:Ch6-Lpoverall} becomes,
\begin{align}
\mathbf{L}_{\mathrm{p}} = \varphi \mathbf{L}_{\mathrm{p},{\gamma}}^{\mathrm{IHP}} + \mathbf{L}_{\mathrm{p}, {\alpha^{\prime}}}. 
\end{align}
 For each crystalline phase, $\mathbf{L}_{\mathrm{p}}$ is the summation over individual shear rate contributions, $\dot{\gamma}^{\alpha}$, of crystallographic slip systems $\alpha$, with the initial slip direction $\mathbf{s}_{0}$ and slip plane normal $\mathbf{n}_{0}$,
\begin{align}
\mathbf{L}_{\mathrm{p},\gamma}=\varphi \sum_{\alpha=1}^{n_{s,\gamma}=3} \dot{\gamma}_{\gamma}^{\alpha} \, \underbrace{\left( \mathbf{s}_{0,\gamma}^{\alpha}\otimes \mathbf{n}_{0,\gamma}^{\alpha} \right)}_{\mathbf{P}_{0,\gamma}^{\alpha}} + \sum_{\alpha=1}^{n_{s,\alpha^\prime}=12/24/48} \dot{\gamma}_{\alpha^{\prime}}^{\alpha} \, \underbrace {\left( \mathbf{s}_{0,\alpha^\prime}^{\alpha}\otimes \mathbf{n}_{0,\alpha^\prime}^{\alpha} \right)}_{\mathbf{P}_{0,\alpha^\prime}^{\alpha}}.
\end{align}
 Here, ${n_s}$ denotes the number of slip systems in lath martensite or the retained austenite film. Note that, since the boundary sliding mechanism is considered as a planar contribution of retained austenite living in the habit plane, only 3 slip systems of the fcc crystal are accounted for, i.e. ${n_{s,\gamma}=3}$. 
 
 It is assumed that the crystallographic slip rate, $\dot{\gamma}$, on a certain slip system is driven by a critical resolved shear stress, $\tau=\left( \mathbf{F}_{\mathrm{e}}^{\mathrm{T}} \cdot \mathbf{F}_{\mathrm{e}} \cdot \mathbf{S}_{\mathrm{e}}\right) : \mathbf{P}_{0} $, and can be described by a visco-plastic rate-dependent relation \citep{hutchinson1976bounds, WIJNEN2021111094},
\begin{align}
\dot{\gamma}^{\alpha}=\dot{\gamma}_0^{\alpha}\left(\dfrac{|\tau^{\alpha}|}{s^{\alpha}}   \right)^{1/m}\mathrm{sign}(\tau^{\alpha}),
\end{align}
in which $s^{\alpha}$ is the slip resistance, $\dot{\gamma}_0^{\alpha}$ denotes the reference slip rate, and $m$ is the rate sensitivity factor.
The kinetics of the slip resistance evolution,
\begin{align}
\dot{s}^{\alpha}= \sum_{\beta=1}^{N_s}h^{\alpha\beta}|\dot{\gamma}^{\beta}|,
\end{align}
are controlled by the hardening modulus, $h^{\alpha\beta}$, evolving due to self hardening of the slip system $\alpha$ and latent hardening induced by other systems, $\beta$, given by \citep{bronkhorst1992polycrystalline},
\begin{align}
h^{\alpha\beta} = h_{0} \left( 1-\dfrac{s^{\alpha}}{s_{\infty}} \right)^{a} (q+(1-q)\delta^{\alpha\beta}).
\end{align}
	Here, $h_{0} $ denotes the reference hardening modulus, $q$ is the latent hardening ratio, and $\delta^{\alpha\beta}$ is the Kronecker delta. 
	The flow resistance $s^{\alpha}$ on $\alpha^{\mathrm{th}}$ slip system evolves from the initial value $s_{0}$ towards the saturation value $s_{\infty}$.
	The model parameters used for ferrite, retained austenite, and lath martensite slip systems are taken from \citep{MARESCA2016198}, and given below in Table \ref{tab:Ch6-PhaseBeh}.  

\begin{table}[ht!]
  \caption{Model parameters used for simulating the response of the phases in the ROI.}
  \centering
  \begin{threeparttable}
    \begin{tabular}{cc@{\qquad}ccc}
        \midrule
        Property & Ferrite & fcc & bcc & Units  \\ \midrule\midrule 
        \makecell{$C11$} & $233.3$ & $ 417$ & $417$ & $GPa$ \\      
        \makecell{$C12$} & $135$ & $ 242.4$ & $ 242.4$ & $GPa$ \\      
        \makecell{$C44$} & $118$ & $211.1$ & $211.1$ & $GPa$ \\      \cmidrule(l r){1-4}
        \makecell{$\dot{\gamma}_{0}$} & $0.001$ & $0.001$ & $0.001$ & $s^{-1}$\\  
        \makecell{$s_{0}$} & $150$ & $270$ & $510$ & $MPa$ \\
        \makecell{$s_{\infty}$} & $250$ & $340$ & $ 2000$ & $MPa$ \\ 
        \makecell{$h_{0}$} & $500$& $250$ & $1500$ & $MPa$ \\      
        \makecell{$n$} & $20$& $20$ & $20$ & -- \\  
        \makecell{$q$} & $1.4$& $1.4$ & $1.4$ & -- \\    
        \makecell{$a$} & $1.32$ & $1.5$ & $1.5$ & -- \\ \midrule\midrule
    \end{tabular}
  \end{threeparttable}
  \label{tab:Ch6-PhaseBeh}
  \end{table}

\section{Applied loading condition}
\label{sec:Ch6-TensionLoad}
To mimic the loading conditions of the experiments, the following deformation gradient rate, $\mathbf{F}$, and first Piola-Kirchhoff, $\mathbf{P}$, are applied in $x$-direction,
\begin{equation}\label{TensionLoad}
\mathbf{\dot{\overline{F}}}=\begin{bmatrix}
\dot{\lambda}  & \ast  & \ast \\
0     & \ast  & \ast \\
0     & 0     & \ast
\end{bmatrix}, \, \mathbf{\overline{P}}=\begin{bmatrix}
\ast  & 0     & 0   \\
\ast  & 0     & 0 \\
\ast  & \ast  & 0
\end{bmatrix}, 
\end{equation}
	where $\dot{\lambda}$ is the stretch rate, and '$\overline{\square}$' denotes that the quantity below is applied on average.
	The symbol '$\ast$' means that the particular component of the tensor is free to evolve. 
	The out-of-plane slip lines appearing on the surface of the sample, shown in \ref{fig:experiments}(b), are the result of the free surface of the sample allowing the planes to slip in that direction without any constraint. 
	This is reflected in the applied periodic boundary conditions by allowing the shear components of the deformation gradient, $F_{12}, F_{13}, F_{23}$, free to evolve in a stress-free condition.

\newpage

\bibliography{/home/ador/Desktop/Papers/4thPaper/FinalVersion/NewVersionFM/Arxiv/main.bib}
\bibliographystyle{apalike}

\end{document}